# *E*-field measurement of pulse line ion accelerator*

WANG Bo(王博)[1] ZENG Rong(曾嵘)[1; 1)] NIU Ben(牛犇)[1] SHEN Xiao-Li(沈晓丽)[1]
SHEN Xiao-Kang(申晓康)[2] CAO Shu-Chun(曹树春)[2] ZHANG Zi-Min(张子民)[2]

[1] State Key Lab of Power Systems, Tsinghua University, Beijing 100084, China
[2] Institute of Modern Physics, Chinese Academy of Sciences, Lanzhou 73000, China

**Abstract**：The *E*-field of pulse line ion accelerator (PLIA) is unique with high frequency (~MHz), large magnitude (~MV/m) and limited measuring space (~cm). The integrated optical *E*-field sensor (IOES) has remarkable advantages and has been used for PLIA *E*-field measurement. Firstly, the transfer function of IOES has been calibrated to ensure the measurement accuracy. And the time-domain response illustrates that the sensor has a fast dynamic performance to effectively follow a 4 ns rising edge. Then, the *E*-field distribution along the axis and near the insulator surface of PLIA was measured, showing that the propagation of *E*-field is almost lossless and *E*-field near the insulation surface is about 1.1 times larger than that along the axis, which is in accordance with the simulation result.

## 1 Introduction

Pulse line ion accelerator (PLIA) is a new kind of heavy ion accelerator in development with research for high energy density physics and warm dense matter. The ion bunch can be accelerated to several tens of MeV, with an accelerating gradient as high as ~MV/m in theory [1, 2]. Compared with a traditional line induction accelerator, the major advantage of PLIA is the lower cost, as well as the axial bunching during the accelerating process. The principle of PLIA was first proposed by the Laurence Berkley National Laboratory, where a prototype has been built [1]. Due to the limited space inside PLIA (cm order), the large *E*-field strength (MV/m order), and the high frequency (MHz order), detection of PLIA accelerating *E*-field is difficult. On the other hand, integrated optical *E*-field sensor (IOES) based on waveguide technology can be applied to measurements of the *E*-field under various environmental conditions, with the permanent attachment of fiber and waveguide. IOES developed by Tsinghua Univ. has a dynamic range from ~kV/m to ~MV/m, a flat frequency response from ~10 Hz to ~100 MHz, and a cm-order dimension, which is competent for PLIA accelerating field measurement [3, 4].

* Supported by the Fund of the National Priority Basic Research of China ( 2011CB209403 ) and National Natural Science Foundation of China under Grant No. 51107063.
1) E-mail: zengrong@tsinghua.edu.cn

As far as we know, only simulation of PLIA accelerating field has been studied [2], and no measurement work has been implemented. The purpose of this paper is to present the measurement of PLIA accelerating field by IOES, which is significant for the optimization of the PLIA device and simulation model.

## 2 Calibration of IOES

IOESs with different dynamic range have been developed [3, 4]. Although the PLIA has an expected accelerating gradient of MV/m order, the Lanzhou Test PLIA used in this study only has an accelerating field of several kV/m [5]. So the IOESs with dipole antenna, which have a maximum measurable *E*-field of 50 kV/m to 300 kV/m were selected for the measurement, and the structure is schematically illustrated in Fig. 1(a). The Mach-Zehnder waveguide is fabricated on the $LiNbO_3$ crystal. The incident linearly polarized laser is equally divided by the first Y branch and propagating along the waveguide arms. With *E*-field in the Y direction, the dipole antenna induces a voltage and the enhanced *E*-field will be generated between the electrodes. *E*-fields imposed on the two waveguide arms have the same magnitude and an opposite direction, due to the push-pull structure of the electrode. The light wave propagating in the two waveguide arms could be modulated by *E*-field according to Pockels effect [6], resulting in a phase difference $\varphi(E) \propto E$. The light waves are combined in the second Y branch, where interference occurs. The modulated optical signal is then sent to the optical receiver and is converted to voltage signal $V_{out}$. The transfer function of the whole measuring system can be expressed as

$$\boldsymbol{V}_{\mathbf{out}} = A[1 + b\cos(\varphi_0 + k\boldsymbol{E})] \quad (1)$$

where $A$ represents the photo-electric conversion coefficient and the transmission loss; $b$ stands for the extinction ratio, determined by the symmetry of the Y branch; $k$ represents the sensor sensitivity; $\varphi_0$ is the optical bias, depending on the intrinsic optical path difference of the waveguide arms. The four parameters $A$, $b$, $k$ and $\varphi_0$ decide the sensor performance, and can be obtained by calibration experiment. The *E*-field strength can be calculated by Eq. (2).

$$\boldsymbol{E} = (\arccos[(\boldsymbol{V}_{\mathbf{out}} - A) / Ab] - \varphi_0) / k \quad (2)$$

The packaged sensor is shown in Fig. 1(b), with a dimension of 7 cm × 1 cm × 1 cm. The integral antenna and electrode are the only metal material in the sensor, with a length of 5 mm, a width of 270 μm and a thickness of 0.5 nm.

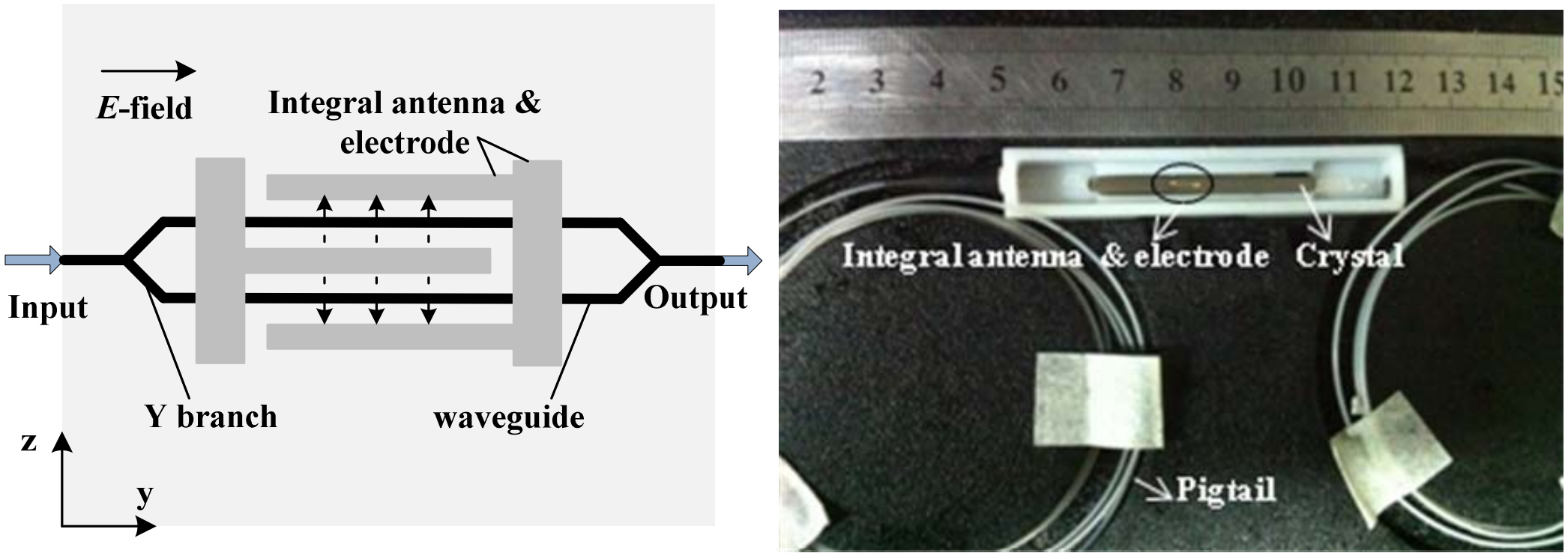

(a) (b)

Fig. 1. (a) Structure diagram of IOES; (b) Sensor after encapsulation.

## 2.1 Transfer function calibration

The sensor transfer function (i.e. the four parameters $A$, $b$, $k$ and $\varphi_0$) can be acquired by calibration under intense $E$-field. The experiment configuration is shown in Fig. 2(a), where the impulse voltage source can generate standard lightning waves with a wave-front time of 1.2 μs and a half-wave time of 50 μs. The parallel electrodes were made according to GB/12720-91, with a separation of $d$=30 cm, and a dimension of 1 m × 1 m. The input voltage $U$ can be obtained by the voltage divider, and the uniform $E$-field generated between the electrodes equals $U/d$, with a relative error less than 0.5 % [7].

Obtaining the input $E$-field and the sensor output simultaneously, the transfer function can be acquired, as shown in Fig. 2(b). The parameters $A$, $b$, $k$ and $\varphi_0$ can be obtained by sinusoidal curve fitting with the experimental data, according to Eq. (1). The root mean squared error RMSE between the fitting result and the measured data is defined as

$$\mathbf{RMSE} = \sqrt{\sum_{i=1}^{n} \frac{1}{n} (V_{oi} - V_{oi}{}^{*})} \qquad (3)$$

and the correlation coefficient $R^2$ is defined as

$$\mathbf{R}^2 = \sqrt{\sum_{i=1}^{n} \frac{1}{n} (V_{oi}{}^{*} - \mu)^2 \Big/ \sum_{i=1}^{n} \frac{1}{n} (V_{oi} - \mu)^2} \qquad (4)$$

where $V_{oi}$ and $V_{oi}$* are the fitting value and the experimental value of the output voltage, $\mu$ is the average value of the output voltage, and $n$ is the number of measuring points. The fitting curve is also shown in Fig. 2(b). The parameters and the results of the regression equation are listed in Table 1. Eq. (1) is an exact description of the input-output character of the measuring system, with $R^2$ equal to 0.9994.

Table 1. The curve fitting result.

| $A$(mV) | $b$ | $k$(m/V) | $\varphi_0$(rad) | RMSE | $R^2$ |
|---|---|---|---|---|---|
| 311.2 | 0.9024 | 0.02419 | 1.63 | 5.281 | 0.9994 |

With the $E$-field strength of 1 kV/m, the measuring system still has a good signal-to-noise ratio. For the convenience of using Eq. (2), the $E$-field should be within the monotonic range of the transfer function (about -50 kV/m to 50 kV/m, according to Fig. 2(b)). So the IOES used in this study has a dynamic range of 1 kV/m to 50 kV/m.

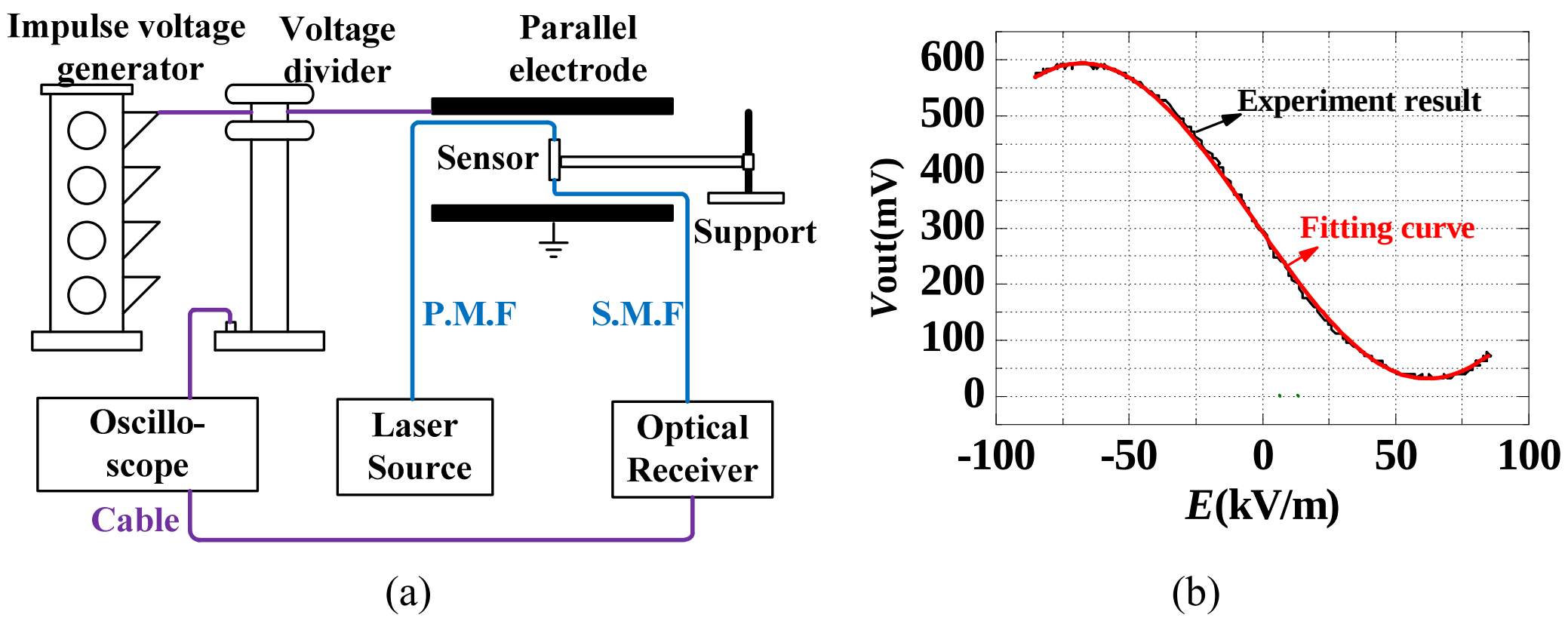

Fig. 2. (a) The experimental setup for transfer function calibration. (b) The measured transfer function and the curve fitting result.

## 2.2 Time-domain response

In order to test the response speed of IOES, an electromagnetic pulse (EMP) with nano-second rise time was generated by an EMP simulator [8], shown in Fig. 3(a). The voltage source is connected to the input end of the gas discharge switch through a protection resistor; and the terminal of the transmission line is a matching load composed of metal film resistors. The voltage $U$ between the transmission lines can be measured by the resistor divider, and the $E$-field in the central point can be calculated by $U/d$ (where $d$ is the separation of the transmission line).

The EMP has a rise time of about 4 ns, and a fall time of about 100 ns. The sensor can respond to the input $E$-field effectively, and the normalized input-output waveforms are shown in Fig. 3(b). The time-domain response illustrates that the sensor has adequate response speed for detecting PLIA field, which has a rise time of about 100 ns. The transfer function could not be obtained through EMP field, since the amplitude is not high enough.

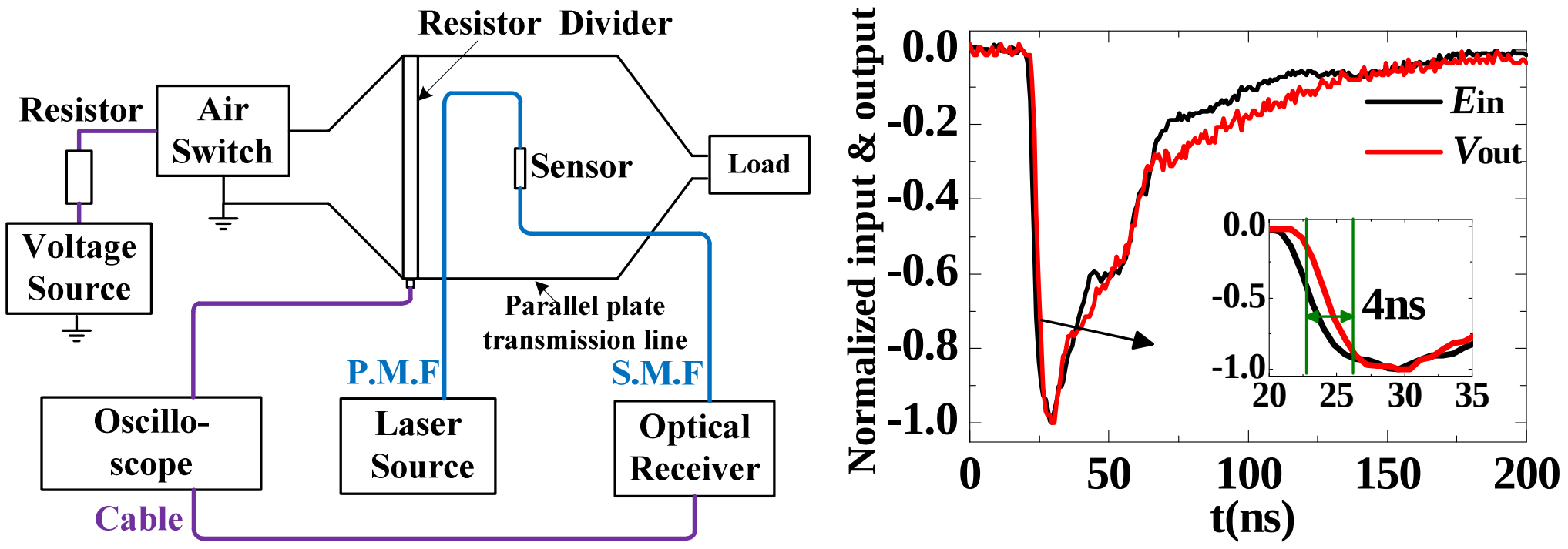

Fig. 3. (a) EMP field generator. (b) Response of EMP field with a rise time about 4 ns.

With a small *E*-field strength, the sensor has an approximate linear input-output characteristic (shown in Fig. 3(b)). Under this circumstance, the scale factor is defined as the peak value of *E* divided by the peak value of $V_{out}$. The scale factor under lightning field is identical with that under EMP field, shown in Table 2. The time-domain results indicate that the sensor has a flat frequency characteristic from ~1 kHz to ~200 MHz, which is consistent with the results of sweep-frequency test [3]. The transfer function calibrated under lightning field can be applied to *E*-field detection with higher frequency.

Table 2. Scale factors under lightning field and EMP field.

| *E*-field type | Rise time | Frequency range | Scale factor |
|---|---|---|---|
| Lightning | 1.2 µs | ~1 kHz to ~1 MHz [9] | 1 mV←→0.205 kV/m |
| EMP | 4 ns | ~10 MHz to ~200 MHz [10] | 1 mV←→0.197 kV/m |

# 3 Experimental measurements

## 3.1 Experimental setup

The Lanzhou Test PLIA is shown in Fig. 4(a), and the inside glass pipe has a length of 90 cm and a diameter of 52 mm [5]. The cross-section dimension of the IOES is 10 mm, which is suitable to measure the PLIA field. The experimental configuration is shown in Fig. 4(b). The sensor can only detect the axial *E*-field of PLIA due to the good directionality, and *E*-field is designated as the axial field in this paper.

The high-frequency and high-voltage pulse source (CPS-10kV-70ns) was connected to the input end of the helix by coaxial cable, and the input waveform was detected by a voltage probe (Tek P6015A). The selected measuring points were 25 cm, 45 cm and 65 cm away from the front end of the glass pipe. The sensor was located on the axis by a plastic holder, to measure the *E*-field along the axis; In addition, the sensor was set on the pipe wall, to detect the *E*-field near the insulation surface (the antenna of the sensor is 5 mm away from the pipe wall).

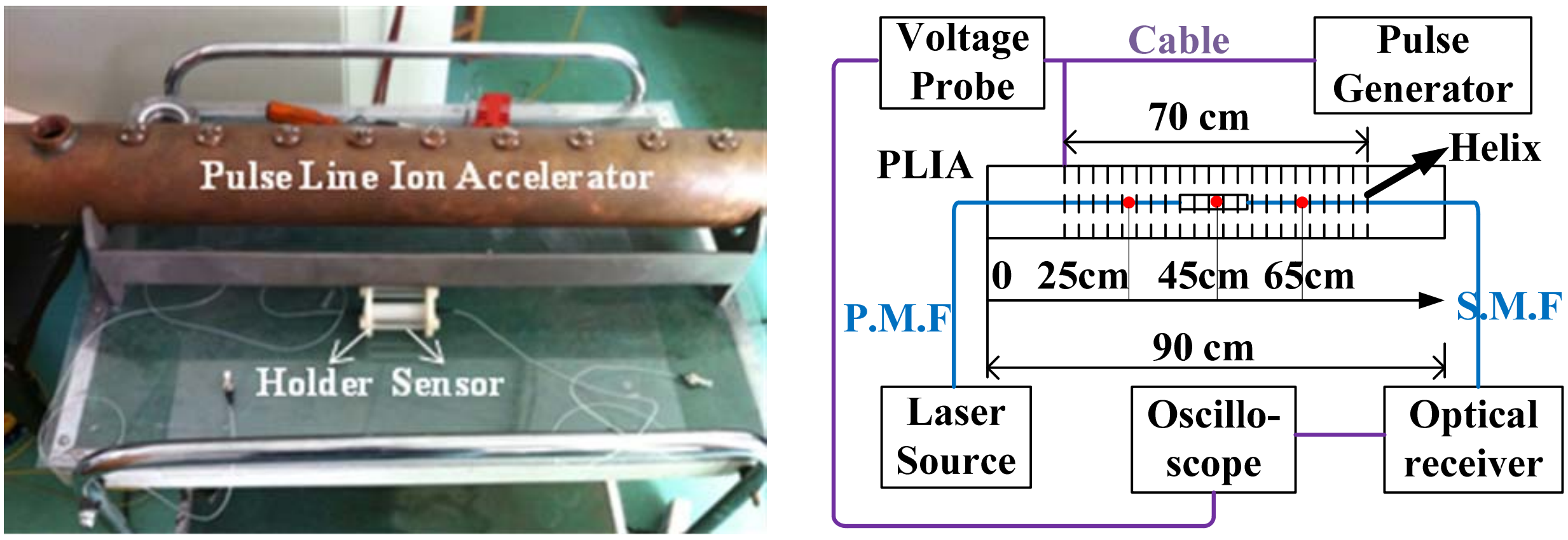

(a) (b)

Fig. 4. (a) Photograph of the PLIA and the IOES. (b) Schematic diagram of the experimental devices.

The input voltage waveform is shown in Fig. 5, which has a good repeatability. The rise time $t_1$ is 80 ns, and the amplitude $V_1$ is adjustable with a range of 0-10 kV. For the convenience of analysis, $V_1$ is normalized to 1 kV during data processing.

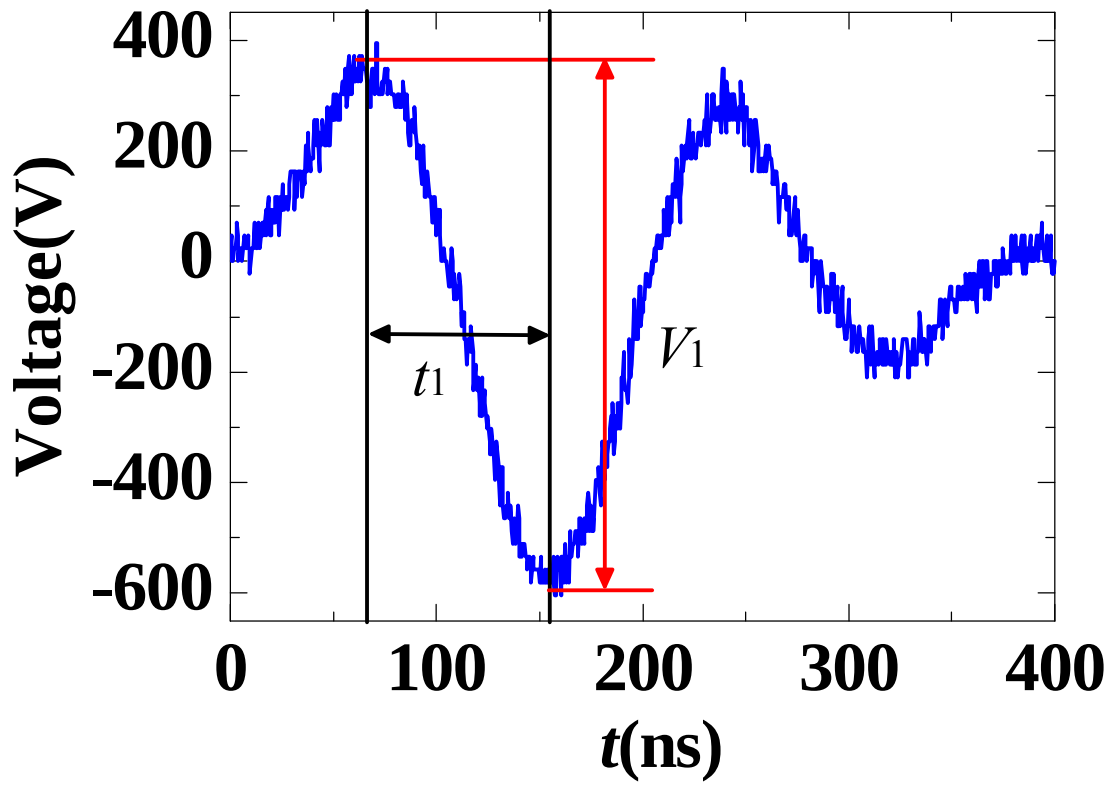

Fig. 5. The input voltage waveform.

### 3.2 Measured result

#### (1) *E*-field along the axis

The *E*-field waveforms along the axis of the PLIA are shown in Fig. 6. The waveforms are very close in different locations, showing that the propagation of the electromagnetic field is nearly lossless. With $V_1$=1 kV, the peak value of the *E*-field along the axis is about 2.9 kV/m.

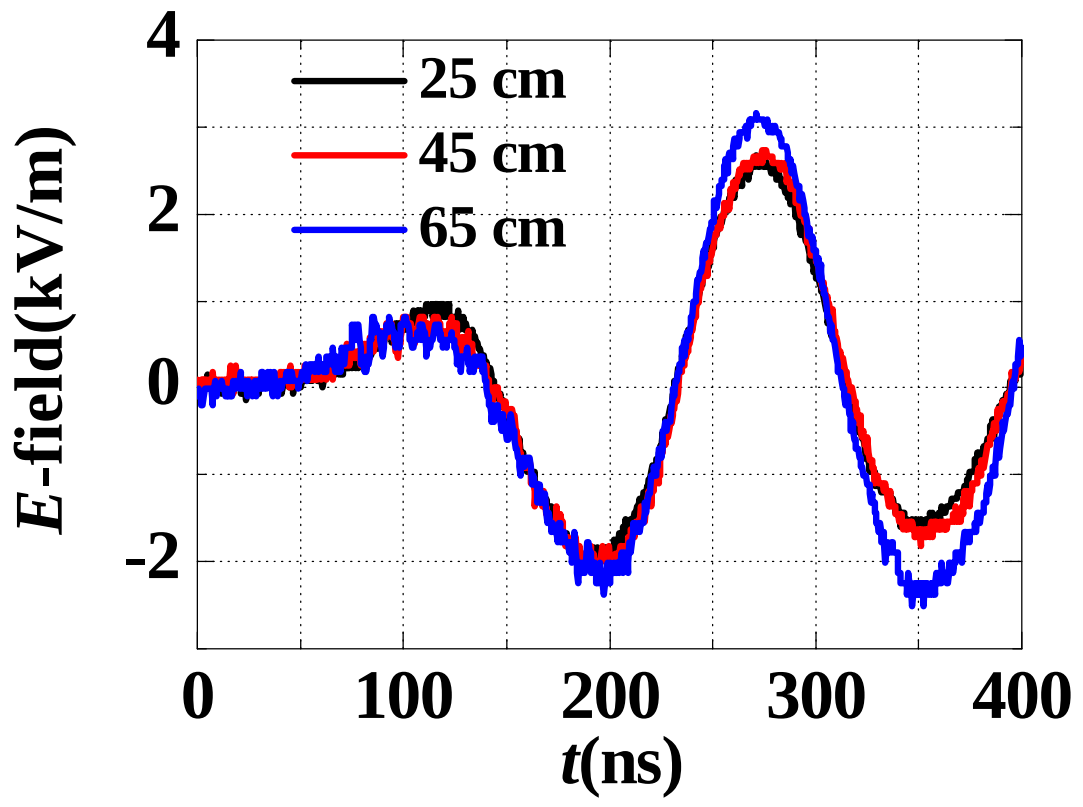

Fig. 6. *E*-field along the axis

**(2) *E*-field around the insulation surface**

The *E*-field waveforms near the insulation surface are shown in Fig. 7, which have a similar distribution characteristic with that on the axis. The waveforms are almost the same in different locations. With $V_1$=1 kV, the peak value of the axial *E*-field around the insulation surface is about 3.3 kV/m.

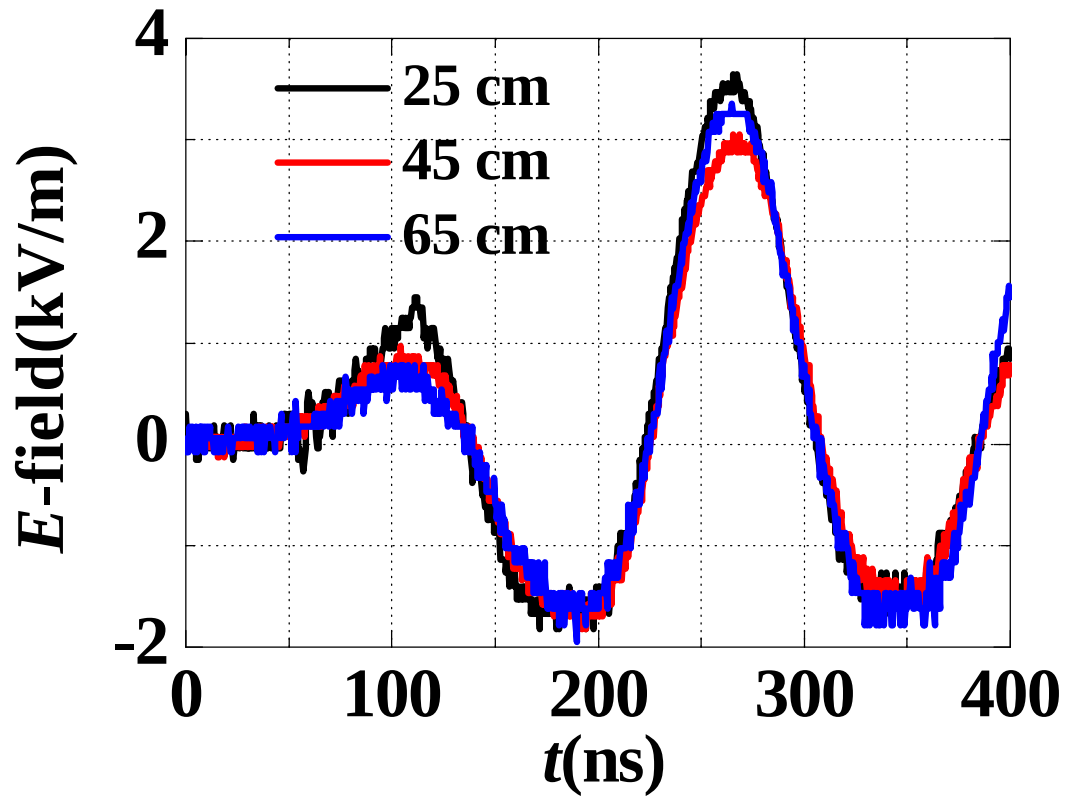

Fig. 7. The distribution axial *E*-field near the pipe wall.

**(3) Comparison between the *E*-field along the axis and that near the insulation surface**

For the measuring point in the center of the PLIA (45 cm away from the pipe end), the *E*-field waveform near the insulation surface is almost the same as that along the axis, but with a 1.1 times larger magnitude.

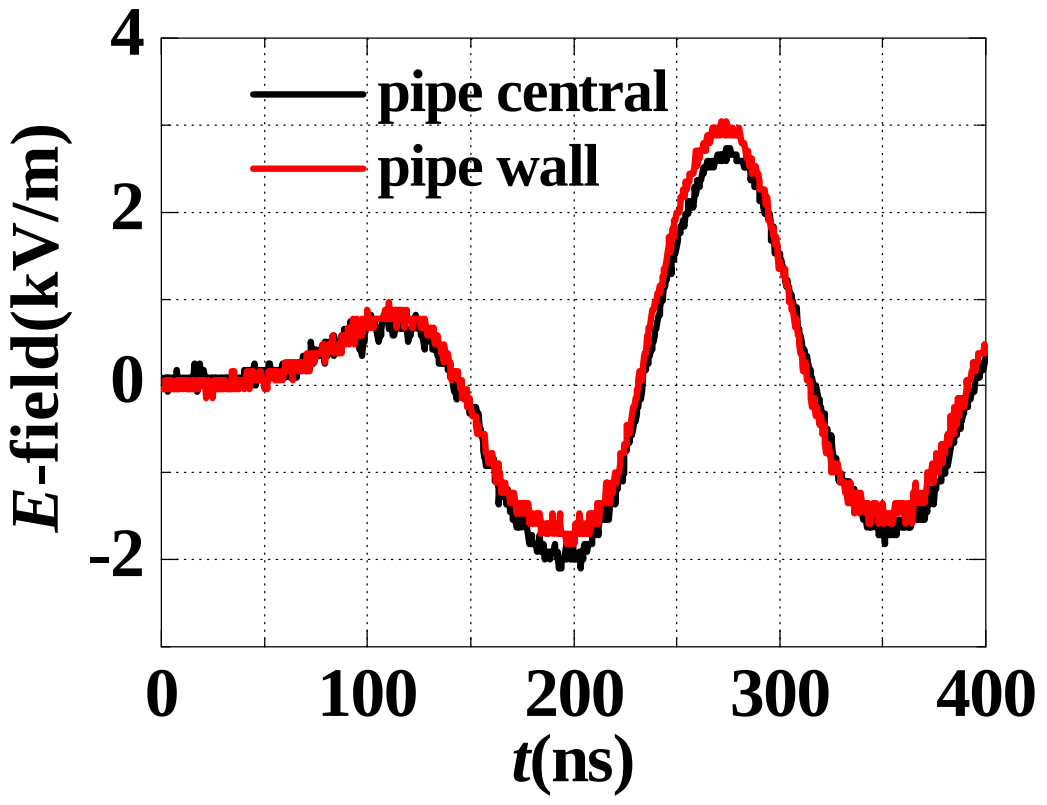

Fig. 8. Comparison between the *E*-field near the insulation surface and that along the axis. The measuring point is in the center of the PLIA.

### 3.3 Comparison between the measurement and simulation

A simulation model of PLIA was built using a general-purpose electromagnetic simulator CST, which is based on the finite integration technique. The voltage shown in Fig. 5 was used as the excitation signal. The simulated waveform is in accordance with the measured one, shown in Fig. 9.

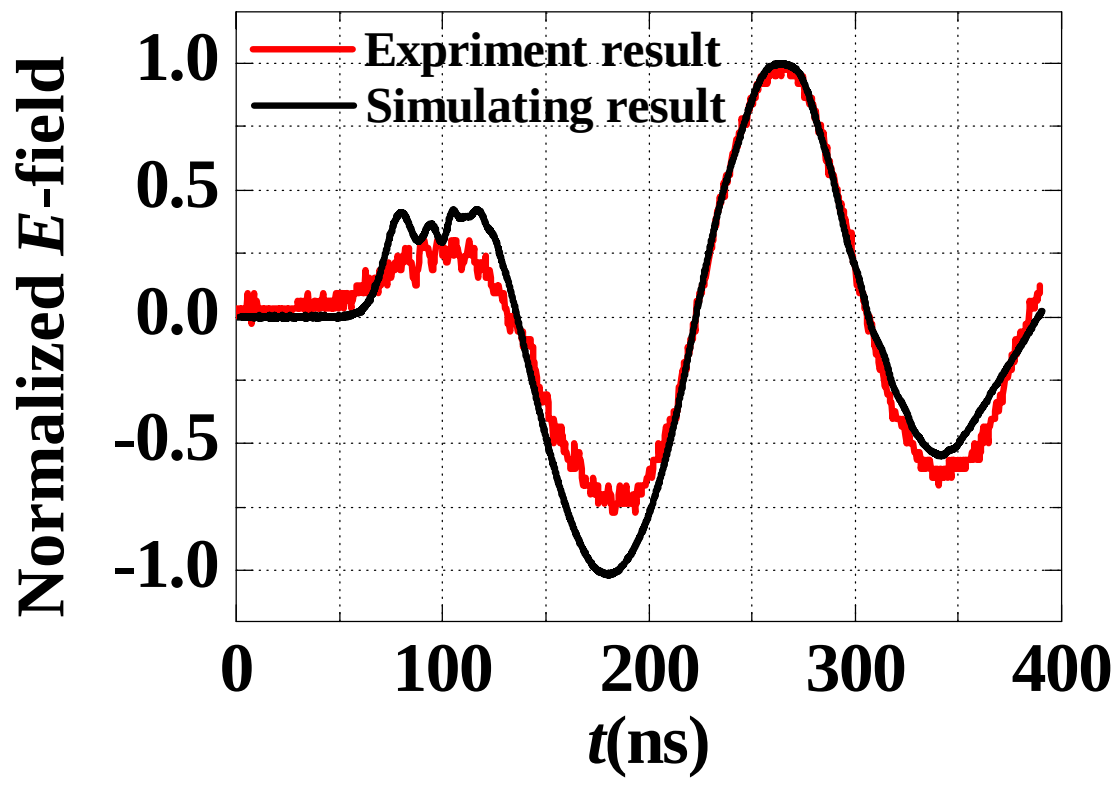

Fig. 9. The normalized simulated result and measured result of the PLIA *E*-field.

The simulated and measured values are shown in Table 3. The simulated peak value is about 2.6 times larger than the measured one. The reason for the difference between the measurement and the simulation may be due to the structural imperfection of the PLIA prototype. The simulation shows that the inductance at the end of the helix has a significant influence on the accelerating field. The helix inductance decreased at both ends due to the reduction of the mutual inductance. A resister string was cascaded to the helix to increase the inductance, which could enhance the accelerating field by 1.85 times in theory [10]. In the PLIA prototype, the resister string may not effectively improve the inductance at the end of the helix, resulting in a lower accelerating field in practice. An improved PLIA with optimized resister string is in construction now.

Both the measurement and simulation show that: 1) the propagation of *E*-field is almost lossless, with a similar value in different positions. 2) *E*-field near the insulation surface is about 1.1 times larger than that along the axis. The measured result could be taken as an important reference in the improvement of the PLIA and the simulation model.

Table 3. The simulated result and measured result with $V_1$=1 kV.

| Position(cm) | Measurement Value(kV/m) | Calculation Value(kV/m) [5] |
|---|---|---|
| Axis(25) | 2.7 | 7.5 |
| Axis(45) | 2.7 | 7.5 |
| Axis(65) | 3.2 | 7.5 |
| Surface(25) | 3.6 | 8.3 |
| Surface(45) | 3.0 | 8.3 |
| Surface(65) | 3.3 | 8.3 |

## 4 Conclusions

The *E*-field of PLIA was measured for the first time.

(a) The IOES is suitable for measuring the *E*-field of PLIA, with a small dimension (7 cm × 1 cm × 1cm), a large dynamic range (1 kV/m-50kV/m), and a fast response speed (less than 4 ns).

(b) The *E*-field distributions along the axis and near the insulation surface were measured, showing that the propagation of *E*-field is almost lossless and the *E*-field near the insulation surface is about 1.1 times larger than that on the axis, which is in accordance with the simulation.

(c) The simulated peak value is about 2.6 times larger than the measured one. The reason for the difference between the measurement and the simulation is probably due to the structural imperfection of the PLIA prototype.